\documentclass[aps,pre,showpacs,twocolumn,amsmath,amssymb,amsfonts,bibnotes,footinbib,a4paper]{revtex4}
\usepackage[dvips]{graphicx}
\usepackage{bm}

\begin{document}

\pacs{05.65.+b,0.5.45.Df,91.30.Px}

\title{Finite driving rate and anisotropy effects in landslide modeling }
\author{E. Piegari, V. Cataudella, R. Di Maio, L. Milano, and M. Nicodemi}
\affiliation{Dipartimento di Scienze Fisiche, Universit\`{a} di
Napoli ``Federico II'', INFM-Coherentia and INFN, Napoli, Italy}

\begin{abstract}
In order to characterize landslide frequency-size distributions and
individuate hazard scenarios and their possible precursors, we
investigate a cellular automaton where the effects of a finite
driving rate and the anisotropy are taken into account. The model is
able to reproduce observed features of landslide events, such as
power-law distributions, as experimentally reported. We analyze the
key role of the driving rate and show that, as it is increased, a
crossover from power-law to non power-law behaviors occurs. Finally,
a systematic investigation of the model on varying its anisotropy
factors is performed and the full diagram of its dynamical behaviors
is presented.
\end{abstract}

\maketitle

\section{Introduction}
Over the last two decades, the evidence of power laws in
frequency-size distributions of several natural hazards such as
earthquakes \cite{ofc}, volcanic eruptions \cite{simkin}, forest
fires \cite{ff,ff2,colloquium} and landslides
\cite{colloquium,dussauge03} has suggested a relationship between
these complex phenomena and self-organized criticality (SOC)
\cite{bak96}. The idea of SOC \cite{bak87}, applied to many media
exhibiting avalanche dynamics \cite{jensen98,turcotte99}, refers to
the tendency of natural systems to self-organize into a critical
state where the distribution of event sizes is represented by a
power law with an exponent $\alpha$, which is universal in the sense
that it is robust to minor changes in the system. Generally, the
nature of a \emph{critical} state is evidenced by the fact that the
size of a disturbance to the system is a poor predictor of the
system response. Let us consider storms as perturbations for natural
slopes. Large storms can produce large avalanches, but also small
storms sometimes can do it. On the other hand, small storms usually
do not produce any avalanche, but also large storms may not cause
any avalanching phenomena. Moreover, avalanches triggered by small
storms can be larger than those triggered by large storms. The
unpredictability of the sizes of such system responses to
incremental perturbations and the observed power-law statistics
could be the exhibition of self-organized critical behavior in most
natural avalanches. However, the idea of understanding power-law
distributions within the framework of SOC is not the only one.
Recently, in order to reproduce the range of the power-law exponents
observed for landslides, some authors have introduced a
two-threshold cellular automaton, which relates landslide dynamics
to the vicinity of a breakdown point rather than to
self-organization \cite{faillettaz}.

In this paper, we report an accurate investigation of a cellular
automaton model, which we have recently proposed to describe
landslide events and, specifically, their frequency-size
distributions \cite{grl}. In particular, we discuss the role of a
finite driving rate and the anisotropy effects in our
non-conservative system. It has been pointed out by several authors
that the driving rate is a parameter that has to be fine tuned to
zero in order to observe criticality \cite{sornette,vespignani,HNJ}.
We notice that the limit of zero driving rate is only attainable in
an ideal chain reaction, therefore finite rates of external drives
are essential ingredients in the analysis of the dynamics of real
avalanche processes. We show that increasing the driving rate the
frequency-size distribution of landslide events evolves continuously
from a power-law (small driving rates) to an exponential (Gaussian)
function (large driving rates). Interestingly, a crossover regime
characterized by a maximum of the distribution at small sizes and a
power-law decay at medium and large sizes is found in the
intermediate range of values of the driving rate for a wide range of
level of conservation. Power-law behaviors are robust even though
their exponents depend on the system parameters (e.g., driving rate
and level of conservation, see below).

Although the critical nature of landslides is not fully assessed and
many authors believe that deviations from power law appear to be
systematic for small landslides data \cite{brardinoni,malamud04},
results from several regional landslide inventories show robust
power-law distributions of medium and large events with a critical
exponent $\alpha \sim 2.5\pm 0.5$ \cite{dussauge03}. The variation
in the exponents of landslide size distributions is larger than in
the other natural hazards that exhibit scale-invariant size
statistics \cite{hergarten03}. Whether this variation of $\alpha$ is
caused by scatter in the data or because different exponents are
associated with different geology, is an important open question,
which we may contribute to address.

The model we analyze describes the evolution of a space and time
dependent factor of safety field. The factor of safety ($FS$) is
defined as the ratio between resisting forces and driving forces. It
is a complicate function of many dynamical variables (pore water
pressure, lithostatic stress, cohesion coefficients, etc.) whose
rate of change is crucial in the characterization of landslide
events.  A landslide event may include a single landslide or many
thousands. We investigate frequency-size distributions of landslide
events by varying the driving rate of the factor of safety. Although
our probability density distributions are lacking of a direct
comparison with frequency-size distributions of real landslides they
reproduce power-law scaling with an exponent very close to the
observed values. Moreover, they allow us to get insight into the
difficult problem of the determination of possible precursors of
future events.

The paper is organized as follows. In the next Section, we present
the model and briefly discuss the differences between our approach
and previous cellular automata models that have been recently
introduced to characterize landslide frequency-size distributions.
In Section III, we report numerical results obtained by a systematic
investigation of the effects of a finite driving rate on the
frequency-size distribution. The values of the exponent of the
power-law decay are given as a function of the driving rate and the
level of conservation. An accurate analysis of the spatial
distribution of the values of the factor of safety by varying the
driving rate provides useful information for quantifying hazard
scenarios of possible avalanche events. In Section IV, we analyze
the role of anisotropic transfer coefficients, which control the
propagation of the instability. We summarize our results in a phase
diagram that shows the location of power-law and non power-law
scaling regions in the anisotropy parameter space. Conclusions are
summarized in Section V.

\section{The Model}
The instability in clays often starts from a small region,
destabilizes the neighborhood and then propagates \cite{bjerrum}.
Such a progressive slope failure recalls the spreading of avalanches
in the fundamental models of SOC. The term self-organized
criticality (SOC) was coined by Bak, Tang and Wiesenfeld to describe
the phenomenon observed in a particular cellular automaton model,
nowadays known as the sandpile model \cite{bak87}. In the original
sandpile model, the system is perturbed externally by a random
addition of sand grains. Once the slope between two contiguous cells
has reached a threshold value, a fixed amount of sand is transferred
to its neighbors generating a chain reaction or avalanche. The
non-cumulative number of avalanches $N_A$ with area $A$ satisfies a
power-law distribution with a critical exponent $\alpha=1$
\cite{kadanoff}, which is much smaller than the values of the
power-law exponents observed for landslides
\cite{colloquium,dussauge03}. Few years later the paper of Bak et
al. \cite{bak87}, Olami, Feder and Christensen (OFC) recognized the
dynamics of earthquakes as a physical realization of self-organized
criticality and introduced a cellular automaton that gives a good
prediction of the Gutenberg-Richter law \cite{ofc}. Such a model,
whose physical background belongs to the Burridge-Knopoff
spring-block model \cite{bk}, is based on a continuous dynamical
variable which increases uniformly through time till reaches a given
threshold and relaxes. This means that the dynamical variable
decreases, while a part of the loss is transferred to the nearest
neighbors. If this transfer causes one of the neighbors to reach the
threshold value, it relaxes too, resulting in a chain reaction. OFC
recognized that the model still exhibits power-law scaling in the
non-conservative regime, even if the power-law exponent strongly
depends on the level of conservation.

In this paper, we investigate the role of a finite driving rate and
of anisotropy in a non-conservative cellular automaton modeling
landslides \cite{grl}. In such a model, we sketch a natural slope by
using a square grid where each site $i$ is characterized by a local
value of the safety factor $FS_i$. In slope stability analysis, the
factor of safety, $FS$, against slip is defined in terms of the
ratio of the maximum shear strength $\tau_{max}$ to the disturbing
shear stress $\tau$
\begin{equation}
FS = \frac{\tau_{max}}{\tau}.
\end{equation}
The limited amount of stress that a site can support is given by the
empirical Mohr-Coulomb failure criterion: $\tau_{max}=c + (\sigma
-u)\tan\phi$, where $\sigma$ is the total normal stress, $u$ is the
pore-fluid pressure, $\phi$ is the angle of internal friction of the
soil and $c$ is the cohesional (non-frictional) component of the
soil strength \cite{terzaghi}. If $FS>1$, resisting forces exceed
driving forces and the slope remains stable. Slope failure starts
when $FS = 1$. Since a natural slope is a complex non-homogeneous
system characterized by the presence of composite diffusion,
dissipative and driving mechanisms acting in the soil (such as those
on the water content), we consider time and site dependent safety
factor $FS_i$ and treat the local inverse factor of safety $e_i =
1/FS_i$ as the non-conserved dynamical variable of our cellular
automata model \cite{grl}.

The long-term driving of the OFC model is, here, replaced by a
dynamical rule which causes the increases of $e_i$  through the time
with a finite driving rate $\nu$: $e_i(t+\Delta t)= e_i(t) +
\nu\Delta t$.
Such a rule allows us to simulate the effect on the factor of safety
of different complex processes which can change the state of stress
of a cell. The model is driven as long as $e_i < 1$ on all sites
$i$. Then, when a site, say $i$, becomes unstable (i.e., exceeds the
threshold, $e_{th} = 1$) it relaxes with its neighbors according to
the rule:
\begin{equation}
e_i \rightarrow 0; \quad \quad e_{nn} \rightarrow e_{nn}+f_{nn}
e_i,\label{relax}
\end{equation}
where $nn$ denotes the nearest neighbors of site $i$ and $f_{nn}$ is
the fraction of $e_i$ toppling on $nn$. This relaxation rule is
considered to be instantaneous compared to the time scale of the
overall drive and lasts until all sites remain below the threshold.
When $e_i$ reaches the threshold value $1$ and relaxes, the fraction
of $e_i$ moving from the site $i$ to its ``downward'' (resp.
``upward'') neighbor on the square grid is $f_d$ (resp. $f_u$), as
$f_l=f_r$ is the fraction to each of its ``left'' and ``right''
neighbors. The transfer parameters $f_{nn}$ are chosen in order to
individuate a privileged transfer direction: we assume $f_u < f_d$
and $f_l =f_r < f_d$. We notice that the model reproduces features
of the OFC model for earthquakes in the limit case $\nu=0$ and
$f_{nn}=f\leq0.25$. A detailed analysis of the model on varying the
transfer coefficients $f_{nn}$ is reported in Sec. IV.

Since many complex dissipative phenomena (such as evaporation
mechanism, volume contractions, etc. \cite{fredlund}) contribute to
a dissipative stress transfer in  gravity-driven failures, we study
the model in the non-conservative case $C=\sum_{nn}f_{nn}<1$, which
makes our approach different from previous ones within the framework
of SOC \cite{hergarten03}. The conservation level, $C$, and the
anisotropy factors, which we consider here to be uniform, are
actually related to local soil properties (e.g., lithostatic,
frictional and cohesional properties), as well as to the local
geometry of the slope (e.g., its morphology). The rate of change of
the inverse factor of safety, $\nu$, induced by the external drive
(e.g., rainfall), in turn related to soil and slope properties,
quantifies how the triggering mechanisms affect the time derivative
of the FS field.

Recently, in order to reproduce the range of the power-law exponents
observed for landslides, several authors have used two-threshold
cellular automata, which relate landslide dynamics to
self-organization \cite{hergarten00} or to the vicinity of a
breakdown point \cite{faillettaz}. In the first approach
\cite{hergarten00}, a time-dependent criterion for stability, with a
not easy interpretation in terms of governing physics, provides a
power-law exponent close to 2 without any tuning. Therefore, this
approach does not explain the observed variability of $\alpha$. In
Ref. \cite{faillettaz}, the range of  $\alpha$ is found by tuning
the maximum value of the ratio between the thresholds of two failure
modes, the shear failure and the slab failure. However, the
frequency-size distribution of avalanches is obtained by counting
only clusters where shear failures have occurred, considering
conservative transfer processes between adjacent cells with a
different number of nearest neighbors. In this paper, the
investigation of our non-conservative cellular automaton is mainly
devoted to the characterization of landslide event dynamics on
varying the driving rate in order to analyze different hazard
scenarios.


\section{The effect of a finite drive on frequency-size distributions}
\begin{figure}[tbp]
\begin{center}
\includegraphics[width=8cm]{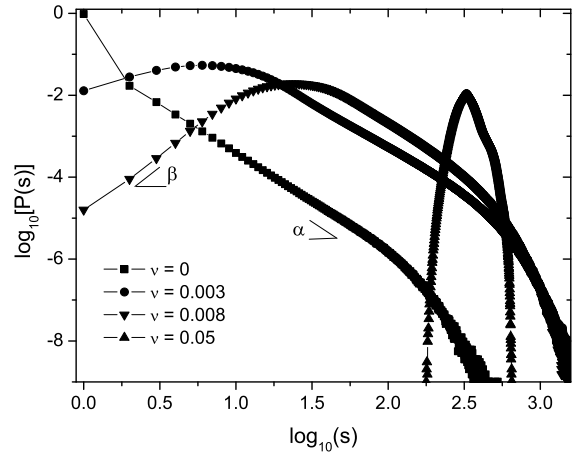}
\end{center}
\caption{Noncumulative frequency-size distributions on a $64\times
64$ grid corresponding to four values of the driving rate. The
logarithm of the normalized number of model events,
$log_{10}[P(s)]$, in which a specified number of different cells,
$s$, become unstable, is given as a function of $log_{10}(s)$. We
show the case $C=0.4$, $f_u/f_d=2/3$ and $f_l/f_d =5/6$.}
\label{scaling}
\end{figure}

Frequency-size distributions give
the number of landslides (events) as a function of their size.
In Fig.\ref{scaling} we show the non-cumulative frequency-size
distributions obtained for different values of the driving rate in
the anisotropic non conservative case $C=0.4$, with $f_u /f_d=2/3$
and $f_l/f_d =5/6$. The curves are obtained for a square lattice of
size $64 \times 64$. We considered both cylindrical (open along the
vertical axis and periodical along the horizontal axis) and open
boundary conditions, which we checked differ in the slopes of the
distribution curves for less than $1\%$.

\begin{figure}
\includegraphics[width=8cm]{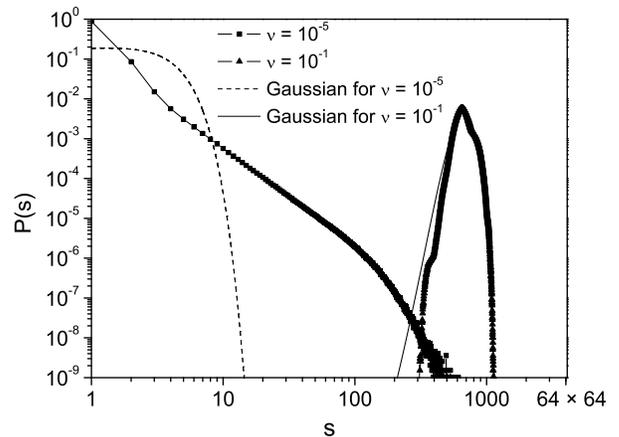}
\caption{Noncumulative frequency-size distributions and
corresponding Gaussian distributions on a $64\times 64$ grid for two
values of the driving rate. Squares and triangles show the
normalized number of model events, $P(s)$, in which a specified
number of different cells, $s$, become unstable as a function of
$s$. The dashed and the solid lines are Gaussian distributions
obtained for the same mean value and the standard deviation of the
frequency-size distributions. We consider the case $C=0.4$,
$f_u/f_d=2/3$ and $f_l/f_d =5/6$.} \label{gaussian}
\end{figure}

In the limit of vanishing driving rate, the distribution of events,
$P(s)$, is similar to that of the two-dimensional isotropic OFC
model for a fixed value of the level of conservation: a power law
characterized by a critical exponent $\alpha$, $P(s)\sim
s^{-\alpha}$, followed by a system finite-size dependent exponential
cutoff \cite{jensen98}. As discussed in Ref. \cite{grl}, by
increasing the driving rate $\nu$, the probability distribution
develops a maximum, which shifts towards larger events with $\nu$.
On the left side of the maximum, a power-law decay, with exponent
$\beta$ (see the Fig. \ref{scaling}), seems to appear for small
landslide sizes. However, the few available data do not allow to
distinguish log-log linear shape and an exponential one
\cite{malamud04}. On the right side of the maximum of the
distribution, the power-law regime remains until, by increasing
$\nu$, the distribution continuously modifies in a bell-shaped
curve. Fig.\ref{gaussian} shows the crossover of the probability
distribution from power-law to Gaussian on increasing the driving
rate.

\begin{figure}
\includegraphics[width=8cm]{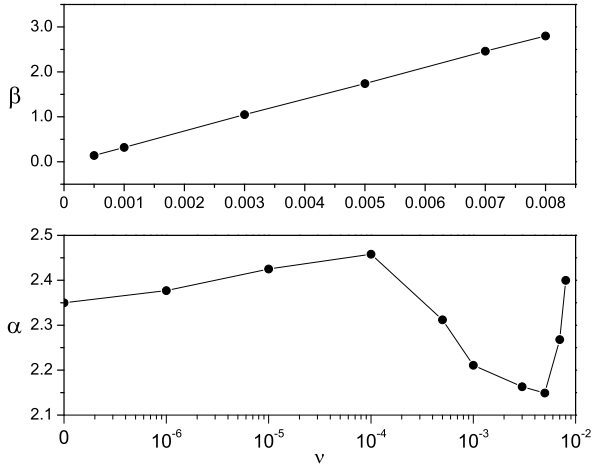}
\caption{Top: Positive power-law exponent $\beta$ as a function of
the driving rate $\nu$. Bottom: Negative power-law exponent $\alpha$
as a function of $\nu$. The values of the exponents are obtained
for $C=0.4$, $f_u/f_d=2/3$ and $f_l/f_d =5/6$. } \label{expodrive}
\end{figure}

\begin{figure}
\begin{center}
\includegraphics[width=8cm]{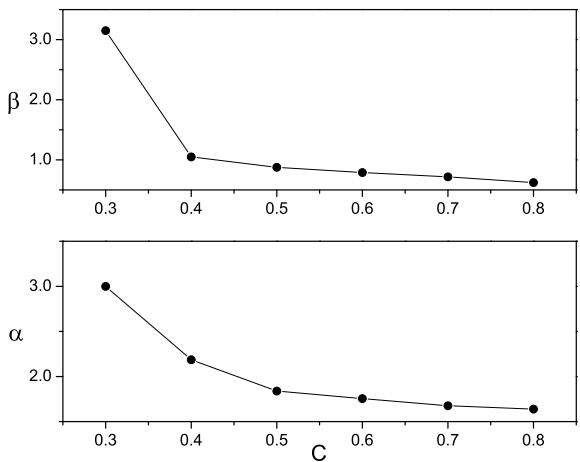}
\end{center}
\caption{Top: Positive power-law exponent $\beta$ as a function of
the level of conservation $C$. Bottom: Negative power-law exponent
$\alpha$ as a function of $C$. The values of the exponents are
obtained for  $\nu=0.003$, $f_u/f_d=2/3$ and $f_l/f_d =5/6$.}
\label{expocons}
\end{figure}

The behavior of the power-law exponents on varying $\nu$ is shown in
Fig.\ref{expodrive}. Interestingly, the values of the power-law
exponent $\alpha$ are very close to those experimentally reported
\cite{colloquium}. As one can see, $\alpha$ increases until the
value of the driving rate sensitively modifies the shape of the
distribution with the appearance of a maximum. We find that the
regime with a non-monotonic frequency-size distribution is robust to
changes in system parameters. In particular, it can be found for
$\nu\in[10^{-4},10^{-2}]$ in the whole range $C\in[0.4,0.8]$.

The values of the exponents $\alpha$ and $\beta$ sensitively change
by varying another important parameter of the model that is the
level of conservation $C$, which represents the non-conservative
redistribution of the load of failing cells. The effect of $C$ at
finite driving rate is comparable to that obtained at $\nu=0$
\cite{ofc}, as shown in Fig.\ref{expocons} (see also Ref.
\cite{grl}).

\begin{figure}[tbp]
\begin{center}
\includegraphics[width=8cm]{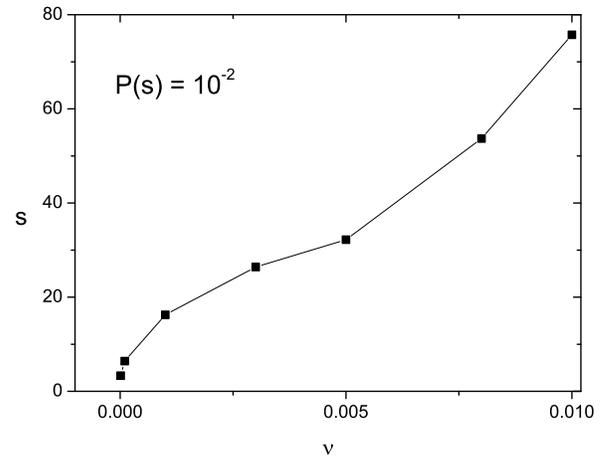}
\hfill
\includegraphics[width=8cm]{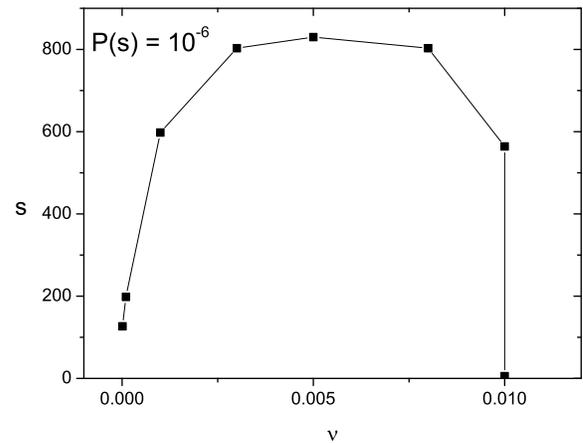}
\end{center}
\caption{ Size, $s$, of equiprobable events (i.e., corresponding to
the same probability of occurrence) as a function of the driving
rate. Top: Sizes of model events with $P(s)=10^{-2}$. Bottom: Sizes
of model events with $P(s)=10^{-6}$. In the top panel, only the
largest sizes of equiprobable events are plotted.
 } \label{ep}
\end{figure}

\begin{figure*}[t]
\caption{(top row) Snapshots of a landslide event of size $s=230$ on
a $64 \times 64$ grid, for four values of the driving rate (from
left to right): $\nu = 10^{-5}$, $\nu = 10^{-3}$, $\nu = 5 \cdot
10^{-3}$ and $\nu = 5 \cdot 10^{-2}$. The 230 black cells are those
that have reached the instability threshold. The simulations are
done in the case $C=0.4$, $f_u/f_d=2/3$ and $f_l/f_d =5/6$. (bottom
row) Snapshots of the factor of safety corresponding to the stable
configurations reached after the avalanches shown in top row. The
values of the factor of safety have been associated to ten levels of
a gray scale from white to black, in order to measure the distance
of a cell from its instability condition: the darker the color, the
farther is the cell from the instability threshold.} \label{map}
\end{figure*}

Let us come back to Fig.\ref{scaling} in order to highlight the role
of the driving rate in the characterization of landslide events.
From the distributions of Fig.\ref{scaling}, we see that large
events can have comparatively high probabilities in the power-law
regime (i.e., at small driving rates $\nu$) with respect to the
Gaussian regime (i.e., larger $\nu$). Such a feature could allow to
reproduce the observation that sometimes mild rainfalls produce
landslides as large as those triggered by intense rainfalls. To
underline the nonlinear behavior of the system with $\nu$, in
Fig.\ref{ep} we plot the sizes of avalanches with the same
probability of occurrence for different values of the driving rate.
Such sizes are the intersections of horizontal straight lines with
the distribution curves of Fig.\ref{scaling}.
When we consider small events (i.e., $P(s)$ as large as $10^{-2}$),
we find that the size of equiprobable events increases with the
driving rate. Thus, for equiprobable events with high probability of
occurrence the system response is essentially linear with the
driving rate. Instead, for large events  (i.e., $P(s)$ as small as
$10^{-6}$), the size of equiprobable events has a maximum as a
function of $\nu$. Thus, it appears that, for a given range of $\nu$
values, the size of events caused by a slow rate of changes of the
factor of safety can be larger than the size of avalanches triggered
by a faster rate. Moreover, an evidence is found for the existence
of the most {\em dangerous} value of $\nu$, for which the size of
the system response has a maximum.

\subsection{Hazards scenarios}
The detection of possible precursors of a landslide event is a
crucial step to achieve hazard reduction. In order to get insight
into this difficult problem, we visualize the structure of a typical
landslide event for different values of the driving rate $\nu$. In
the top row of Fig.\ref{map}, we report on the $64\times 64$ grid a
typical avalanche of size $s=230$ on increasing the value of $\nu$
from left to right. The $230$ black cells are those that have
reached the instability threshold. As we can see, compact landslides
are the characteristic response of a system governed by power-law
statistics as it happens at small $\nu$ (also when a maximum in the
frequency-size distribution develops). Such a response is typical of
systems with SOC behavior \cite{pietronero}. By increasing the
driving rate, compact clusters survive until power-law regime
disappears. As the system enters the non power-law regime, the
relevance of domino effects drastically drops and landslide events
are characterized by many tiny independent clusters.

In the bottom row of Fig.\ref{map}, we show the distribution of the
factor of safety $FS_i =1/e_i$ for the cases corresponding to the
upper panels \cite{nota1}. The distribution on the grid of the
values shows how the spatial correlations in the system crucially
affect landslide structures. In order to measure the distance of a
cell from its instability condition and to visualize the correlated
areas (regions with similar values of $FS_i$), the values of $FS_i$
have been associated to ten levels of a gray scale from white to
black: the darker the color, the farther is the cell from the
instability threshold. In the snapshots of Fig.\ref{map}, it is
possible to recognize as dark areas the avalanches shown in the
corresponding upper grids. In particular, the dark areas typically
are related to previous landslide events, whereas the lighter areas
indicate regions of future events. We notice that in the power law
regime (i.e. small $\nu$) even a very small perturbation (say, a
drop of water) at one single point can trigger huge system
responses. Instead, in the non power-law regime (i.e. large $\nu$)
large-scale correlations are absent; here large events trivially
occur just because the strong external driving rate makes likely
that many cells simultaneously approach the instability threshold.
Thus, the detection of patterns of correlated domains in
investigated areas results to be a crucial tool to identify the
response of the system to perturbations, i.e., to hazard assessment.

It is worth noticing that the average value of the factor of safety
on the grid cells $<FS>$ and its fluctuation $<\Delta FS^2>$ are
very similar in the four cases showed in Fig.\ref{map}, encompassing
a broad spectrum of $\nu$ values. Interestingly, the probability
distribution of $FS$ on the grid sites, independently of the driving
rate, is well approximated by a Gaussian distribution. This suggests
that a measure of just an average safety factor on the investigated
area could provide only a very partial information about the
statistics governing the considered landslide events.



\section{Anisotropy effects on frequency-size distributions}

In the previous sections, we have investigated the properties of the
model on varying the driving rate $\nu$ at fixed values of the
anisotropic ratios $f_u/f_d$ and $f_l/f_d=f_r/f_d$. Such parameters
control how the instability propagates downward and, therefore, they
are complicate functions of the topography and geology of a specific
area. In this Section, we are interested in the analysis of the
model when such transfer coefficients vary.


It is well-known that the one-dimensional version of the sandpile
and OFC models are characterized by non power-law scaling
\cite{jensen98}. Thus, we expect that, for small values of the ratio
$f_l/f_d$, the frequency-size distribution does not show power-law
behavior. Viceversa, for $f_u/f_d\sim 1$ and $f_l/f_d\sim 1$, we
expect to have power-law distributions, as in the OFC model, which
corresponds in our model to the limit $f=f_u=f_l=f_r=f_d$ and
$\nu=0$.

\begin{figure}
\includegraphics[width=8cm]{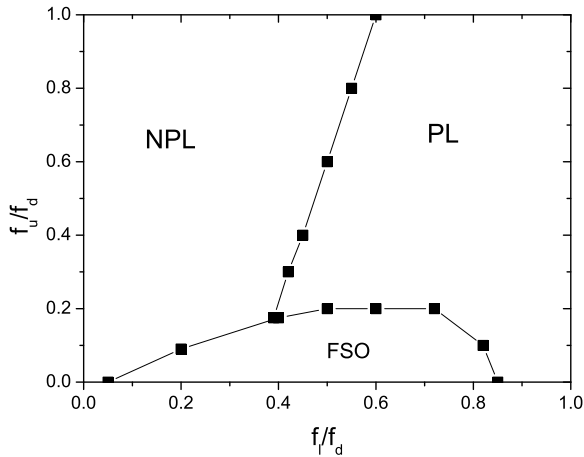}
\caption{Phase diagram in the $f_u/f_d$ vs. $f_l/f_d$ plane. The
lines divide the parameter space in three regions: a power-law
scaling region (PL), a non power-law scaling region (NPL) and a
finite-size oscillation region (FSO) where peaks of the distribution
commensurate with the size of the grid appear. The results are
obtained for $\nu=10^{-4}$ and $C=0.4$.} \label{phd}
\end{figure}

The diagram of Fig.\ref{phd} summarizes the different regimes found
in our simulations at fixed values of the driving rate $\nu=10^{-4}$
and the level of conservation $C=0.4$. We find that, on varying the
anisotropic ratios, the parameter space is divided in three regions:
\emph{i)} a power-law region (PL) characterized by power-law
frequency-size distributions for large values of the anisotropic
ratios, \emph{ii)} a non power-law region (NPL) for the whole range
of  values of the anisotropic ratio $f_u/f_d$ (which controls the
redistribution of load in the vertical direction) and small values
of $f_l/f_d$, \emph{iii)} a finite-size oscillation region (FSO)
where the frequency-size distribution is characterized by periodic
peaks, which appear for integer multiples of the grid size $L=64$.

As expected, we find power-law and non power-law behaviors  for
large and small values of the anisotropic ratios, respectively. In
particular, we find that the value of the critical exponent $\alpha$
of the power-law distribution slightly increases with $f_u/f_d$ for
a fixed value of $f_l/f_d$ and decreases with $f_l/f_d$ for a fixed
value of $f_u/f_d$. However, the changes in $\alpha$ are negligible.

\begin{figure}
\includegraphics[width=8cm]{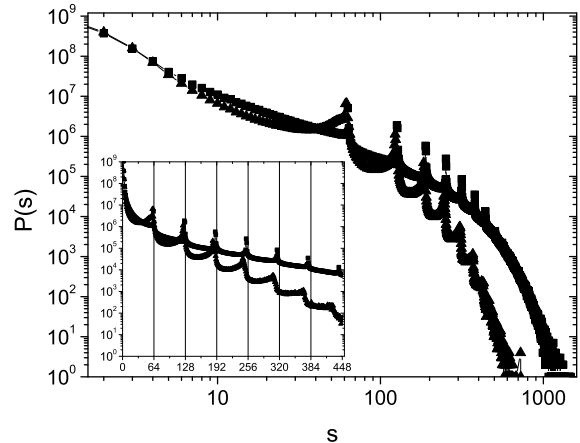}
\caption{Noncumulative frequency-size distributions on a $64\times
64$ grid for $f_u/f_d=0$ and $f_l/f_d=0.3$ (triangles) and
$f_l/f_d=0.55$ (squares). The inset shows the distribution curves in
a log-linear scale for the range of values  where commensurate peaks
are observed. The vertical lines mark multiples of the system size.
The results are obtained for $\nu=10^{-4}$ and $C=0.4$.}
\label{peaks}
\end{figure}

It is worth noticing that, even if the ratio $f_l/f_d$ is quite
large, we find that the frequency-size distribution does not show a
power law decay when the value of $f_u/f_d$ is small, (see
Fig.\ref{phd}). Indeed, the probability distribution develops a
finite number of peaks which are commensurate with the size of the
grid. Fig.\ref{peaks} shows the frequency-size distribution for two
different values of the anisotropic ratios in the FSO region of the
phase diagram. By increasing $f_l/f_d$, the peaks of the
distributions turn down as long as they disappear. We m that several
peaks in the distribution of avalanche sizes are obtained in Ref.
\cite{corral}. The authors vary the convexity of the driving
$\gamma$ and the level of conservation $\epsilon$ in the isotropic
case: $C=1$ when $\epsilon=0.25$. They find an intermediate region
of the phase-diagram $\epsilon$ vs. $\gamma$ where the
frequency-size distribution is characterized by peaks that scale
with different powers of the system size $L$. Fixing the convexity
of the driving $\gamma$ ($\gamma =0$ for a uniform driving), for
large values of the level of conservation, $C\geq 0.6$, the peaks
disappear and they get a power-law decay with an exponential cutoff.
Conversely, we find commensurate peaks in the FSO region in the
whole range $C\in[0.4,0.8]$.

\begin{figure}[tbp]
\includegraphics[width=8cm]{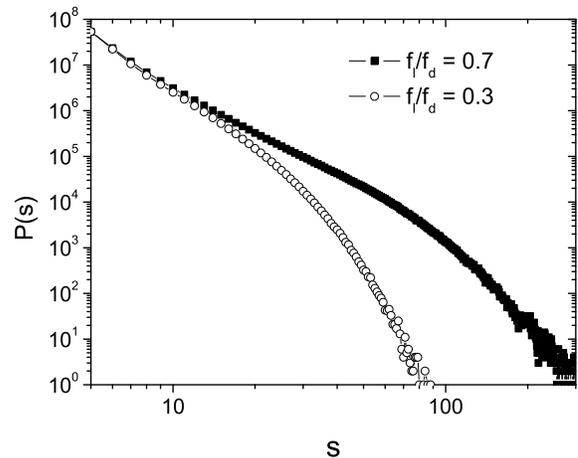}
\caption{Noncumulative frequency-size distributions on a $64\times
64$ grid for $f_u/f_d=1$ and $f_l/f_d=0.3$ (open dots) and
$f_l/f_d=0.7$ (squares). The results are obtained for $\nu=10^{-4}$
and $C=0.4$.} \label{comparison}
\end{figure}

We notice that an analysis of the anisotropic case of the OFC model
is made in Ref.\cite{christensen92} where the authors introduce only
two transfer coefficients $\alpha_1=f_l=f_r$ and $\alpha_2=f_u=f_d$
and control the degree of anisotropy by changing the ratio
$\alpha_1/\alpha_2$, while keeping the level of conservation
constant. They find that the anisotropy has almost no effect on the
power-law exponent while the scaling exponent, expressing how the
finite-size cutoff scales with the system size, changes continuously
from a two-dimensional to a one dimensional scaling of the
avalanches \cite{christensen92}. Varying the anisotropic ratio
$\alpha_1/\alpha_2$ in the range $[0,1]$ is equivalent to consider
the straight line $f_u/f_d=1$ in the phase diagram  of
Fig.\ref{phd}. As in Ref. \cite{christensen92}, we find that on
moving along the line $f_u/f_d=1$, the changes in the power-law
exponent are negligible. However, differently  Ref.
\cite{christensen92}, we find a crossover in the frequency-size
distribution behavior from power-law to non power-law (see Fig.
\ref{comparison}). We attribute such a different result to the
finite driving rate.

In conclusion, our analysis shows that only a finite range of values
of the anisotropic transfer coefficients can supply power-law
distributions. This characterization provides insight into the
difficult determination of the complex and non-linear transfer
processes that occur in a landslide event.

\section{Concluding Remarks}

Explanation of the power-law statistics for landslides is a major
challenge, both from a theoretical point of view as well as for
hazard assessment. In order to characterize frequency-size
distributions of landslide events, we have investigated a
continuously driven anisotropic cellular automaton based on a
dissipative factor of safety field. We have found that the value of
the driving rate, which describes the variation rate of the factor
of safety due to external perturbations, has a crucial role in
determining landslide statistics. In particular, we have shown that,
as the driving rate increases, the frequency-size distribution
continuously changes from power-law to gaussian shapes, offering the
possibility to explain the observed rollover of the data for small
landslides. The values of the calculated power-law exponents are in
good agreement with the observed values. Moreover, the analysis of
the model on varying the driving rate suggests the determination of
correlated spatial domains of the factor of safety as a useful tool
to quantify the severity of future landslide events.

As concerns the effects of anisotropic transfer coefficients, which
control the non-conservative redistribution of the load of failing
cells, we have found that the power-law behavior of the
frequency-size distribution is a feature of the model only in a
limited region of the anisotropy parameter space.

\begin{acknowledgments}
E. Piegari wishes to thank A. Avella for stimulating discussions and
a very friendly collaboration. This work was supported by MIUR-PRIN
2002/FIRB 2002, SAM, CRdC-AMRA, INFM-PCI, EU MRTN-CT-2003-504712.
\end{acknowledgments}


\begin{references}
\bibitem{ofc}Z. Olami, H. J. S. Feder, K. Christensen, Phys. Rev. Lett.
{\bf 68}, 1244 (1992).

\bibitem{simkin}T. Simkin, Annu. Rev. Earth Planet. Sci.
{\bf 21}, 427 (1993).

\bibitem{ff}P. Bak, K. Chen, C. Tang, Phys. Lett. A. {\bf 147}, 297 (1992).

\bibitem{ff2} R. Pastor-Satorras, and A. Vespignani,
Phys. Rev. E {\bf 61}, 4854 (2000).

\bibitem{colloquium}D. L. Turcotte, B. D. Malamud, F. Guzzetti,
P. Reichenbach, Proc. Natl. Acad. Sci. U.S.A. {\bf 99}, 2530
(2002).

\bibitem{dussauge03}C. Dussauge, J.R. Grasso, and A
Helmstetter, J. Geophys. Res. {\bf 108}, 2286 (2003).

\bibitem{bak96}P. Bak, \emph{How Nature Works - The Science of Self-Organized
Criticality}, (Copernicus, Springer-Verlag, New York, 1996).

\bibitem{bak87}P. Bak, C. Tang and K. Wiesenfeld, Phys. Rev. Lett.
{\bf 59}, 381 (1987); Phys. Rev. A   {\bf 38}, 364 (1988).

\bibitem{jensen98}H. J. Jensen, {\em ``Self-Organized Criticality: emergent
complex behavior in physical and biological systems''} (Cambridge
University Press, Cambridge, 1998).

\bibitem{turcotte99}D. L. Turcotte, Rep. Prog. Phys.
{\bf 62}, 1377 (1999).

\bibitem{faillettaz}J. Failletaz, F. Louchet, J.R. Grasso, Phys. Rev. Lett.
{\bf 93}, 208001 (2004).

\bibitem{grl} E. Piegari, V. Cataudella, R. Di Maio, L. Milano,
M. Nicodemi, Geophys. Res. Lett. in press.

\bibitem{HNJ} D. Hamon, M. Nicodemi and H.J. Jensen,
Astronomy\&Astrophysics {\bf 387}, 326 (2002).

\bibitem{vespignani}R. Dickman, A. Vespignani, S. Zapperi, Phys. Rev. E
{\bf 57}, 5095 (1998).

\bibitem{sornette}D. Sornette, \emph{Critical Phenomena in Natural Sciences,
Chaos, Fractals, Self-organization and Disorder: Concepts and
Tools}, (Springer Series in Synergetics, Heidelberg, 2004).

\bibitem{brardinoni} F. Brardinoni, M. Church,
Earth Surf. Process. Landforms {\bf 29}, 115 (2004).

\bibitem{malamud04} B. D. Malamud, D. L. Turcotte, F. Guzzetti, P. Reichenbach,
Earth Surf. Process. Landforms {\bf 29}, 687
(2004).

\bibitem{hergarten03}S. Hergarten, Natural Hazards and Earth
System Sciences {\bf 3}, 505 (2003).

\bibitem{bjerrum}L. Bjerrum,  J. Soil Mech. Fdns. Div. Am. Soc.
Civ. Engnrs. {\bf 93}, 3 1967.


\bibitem{kadanoff} L. P. Kadanoff, S. R. Nagel, L. Wu, S. M. Zhou,
Phys. Rev. A  {\bf 39}, 6524 (1989).

\bibitem{bk}R. Burridge and L. Knopoff, Bull. Seismol. Soc. Am.
 {\bf 57}, 341 (1967).

\bibitem{terzaghi}K. Terzaghi, Geothecnique
{\bf 12}, 251 (1962).

\bibitem{fredlund}D. G.  Fredlund, H. Rahardjo,
{\em ``Soil Mechanics for Unsatured Soils''} (Wiley-Interscience,
New York, 1993).

\bibitem{hergarten00}S. Hertgarten and H. J. Neugebauer, Phys. Rev. E
{\bf 61}, 2382 (2000).

\bibitem{nota1} In order to visualize the $FS_i$ distribution,
we introduced a lower threshold $e_{min}=10^{-3}$ and checked that
the results do not change.

\bibitem{pietronero}L. Pietronero, W. R. Schneider, Phys. Rev. Lett.
{\bf 66}, 2336 (1991).

\bibitem{christensen92}K. Christensen and Z. Olami, Phys. Rev. A
{\bf 46}, 1829 (1992)

\bibitem{corral}A. Corral, C. J. Perez, A. Diaz-Guilera, A. Arenas, Phys.
Rev. Lett. {\bf 74}, 118 (1995)


\end{references}
\end{document}